\documentclass[12pt]{article}
\begin{document}
\title{Macroscopic Reality in Quantum Mechanics;\\ Origin and Dissipation}%

\author{{Kentaro Urasaki}%
\footnote{E-mail address: urasaki@f6.dion.ne.jp}
\footnote{Kamiigusa Institute of Science, Suginami, Tokyo, Japan}}\date{}%

\maketitle
\begin{abstract}
We study the connection between dissipation and reality in 
macroscopic quantum systems. 
We present the following scenario; 
if we consider the dynamics of a `partial' wave function, 
the dissipation is represented as a nonlocal term 
and it causes destructive interference 
to suppress the quantum fluctuation. 
Using the variational method, 
we confirm that this dissipation term is a reasonable extension 
of the standard (Schr\"odinger) description 
for isolated systems, from which we also derive the classical action. 
Consequently, in macroscopic systems, 
the states whose time-integrated dissipation takes an extreme value come true. 
This description, which is consistent with our sense of reality, 
coexists with the usual linear-time-dependent description. 
\end{abstract}

\section{Introduction}
\subsection{Motivation}
Early in the 20th century 
classical mechanics was abandoned 
and giving way to quantum theory 
on the simplest and the most fundamental systems, 
i.e. photons, electrons and atoms. 
After that, although 
the accumulation of diverse experimental 
results has made, 
we have never encountered its limitation. 
Far from it, its foundation as 
a fundamental theory of physics is being more
solidified.

The essence of quantum theory is often 
expressed as noncommutativity or waviness. 
In non-relativistic region, 
this leads quantum mechanics and 
we have some equivalent but different styles 
of mathematical representations. 
In addition, since each of these representations has the ability
to give {\it autonomous} models for various situations 
and both the number of elements and the expanse of the system 
do not restrict these models themselves, 
we recognize that there is an obvious {\it continuity} 
between micro- and macroscopic systems. 
In fact, 
this continuity has brought 
great success on 
our understanding of 
the material world and 
applications. 
Therefore it is natural 
that we should expect quantum mechanics 
to make radical change in everyone's worldview. 

Despite the expectation, 
something seems to limit the applicability 
of quantum concepts or the continuity of 
the description. 
Our daily experience is itself a common example. 
One of the mathematical expressions, 
so-called the Schr\"odinger equation, 
declares that the state of a system 
is a kind of linear wave. 
Therefore, 
in the same way as all the other linear waves, 
any superposition of states can be allowed. 
In fact, this property 
is absolutely important to understand 
the microscopic origin of our material world. 
However, when we extend it to macroscopic phenomena, 
it is sharply confronted 
with our everyday experience; 
we never observe directly such wavy features in 
our surroundings. 
On the other hand, classical mechanics holds 
with amazing accuracy, which can't be expected from 
quantum mechanics. 
Therefore, it occurs to us that quantum features, 
i.e. fluctuation or waviness, seem to be 
perfectly suppressed for some reason.

On the other hand, we also encounter vivid examples in laboratories. 
For example, on a fluorescent screen or 
in a bubble chamber, 
even a single elementary `particle' 
isn't observed as a wave, i.e. a superposition of position states, 
but like a material point. 
Although it is contradictory to the Schr\"odinger equation, 
we thus cannot abandon 
the old-fashioned statement that 
an elementary particle 
is a substance without volume. 
Not only in a position of particle, but in any physical quantity 
we always fail to observe a quantum superposition or waviness. 
The simple question occurs to us; does this mean that the {\it watch} intervenes 
the mathematical description in which the whole world should be
autonomously evolving with time? 
This question is regarded to be reasonable and often called the measurement problem, 
since the gap becomes acute 
when we intend to observe microscopic properties directly. 
Although many researches (for example in \cite{Wheeler1983}) have already been made, 
it stays unsolved 
to explain the origin of the descriptive gap 
between quantum mechanics and our direct experience. 

In contrast to its appearance, 
it is difficult to resolve the above problems into a well-defined style 
since quantum mechanics seems to 
provide an autonomous model for any kind of situation\cite{Neumann1955}. 
In this paper, we particularly focus on the origin of 
the classicality and the reality in macroscopic systems.%
\footnote{%
We use the word `classicality' to refer to 
the fact that the physical quantities behave like those of classical mechanics 
and `macroscopic reality' to refer to our sense of reality, 
which is based on both the classicality and 
the stability of macroscopic systems.}

\subsection{Strategy}
Throughout our study, 
we consider the meaning of the fact that a system 
can be described by a single wave function. 

In \S 2.1, we consider to divide the system into two parts. 
Then we find that the integration of 
the external degrees of freedom naturally leads 
the nonlocal dissipation term. 
In \S 2.2, the dynamics of the 
disappearance of superposition states is shown, where 
we find the conditions for the stability of partial systems. 
It is also shown that 
two descriptions coexist; 
one is that a whole system evolves linearly with time, and 
the other is that partial systems are interacting each other. 
In \S 3.1, it is verified that 
the dissipation term is the natural extension  
of the standard Schr\"odinger equation. 
In \S 3.2, we obtain the conservation laws for physical quantities and 
the classical action.

\section{Relation between macroscopic reality and dissipation}
Usually interacting degrees of freedom are mixed. 
In some cases, however, the wave function of a partial system has an important role. 
For example, a Schr\"odinger equation with an external field is nothing more than 
that we obtain by fixing external degrees of freedom and integrating them. 
Let us consider the conditions that make it possible 
to describe a partial system with a single wave function. 
Then it will be confirmed that 
a nonlocal term is added to the Schr\"odinger equation 
when the system dynamically interacts with outside. 

\subsection{Schr\"odinger equation for a partial system}
In this subsection, we make a preliminary discussion 
about the contact of macroscopic quantum systems. 

Let us start from the familiar Schr\"odinger equation 
satisfied by the wave function of 
a whole system%
\footnote{%
In this paper, 
we neglect spin and quantum statistics for 
simplicity, and take account of only Coulomb interaction.}; 
\begin{eqnarray}\label{Schroedinger-eq}
[-i\hbar\frac{\partial}{\partial t}+\hat{H}]\Phi_0=0, 
\end{eqnarray}
where 
\begin{eqnarray}
\hat{H}=-\sum_{1\le i\le N}\frac{\hbar^2}{2m_i}\nabla_i^2+\sum_{1\le i<j\le N}\frac{q_iq_j}{|{\bf r}_i-{\bf r}_j|}.
\end{eqnarray}
To survey the contact of two systems we try 
an approximate solution, $\Phi=\varphi({\bf r})\Psi({\bf R}).$%
\footnote{$\varphi({\bf r}, t)=\varphi({\bf r}_1,\cdots,{\bf r}_m, t), \Psi({\bf R}, t)=\Psi({\bf r}_{m+1},\cdots,{\bf r}_N, t).$}
Each of the wave function is normalized. 
We can also separate the Hamiltonian as 
$\hat{H}=\hat{h}_\varphi({\bf r})+\hat{h}_{int}+\hat{h}_\Psi({\bf R}),$ 
where 
\begin{eqnarray}
&&\hat{h}_\varphi({\bf r})=-\sum_{1\le i\le m}\frac{\hbar^2}{2m_i}\nabla_i^2+\sum_{1\le i<j\le m}\frac{q_iq_j}{|{\bf r}_i-{\bf r}_j|},\\ 
&&\hat{h}_{int}=\sum_{1\le i\le m<j\le N}\frac{q_iq_j}{|{\bf r}_i-{\bf r}_j|}.
\end{eqnarray}
This is a kind of mean field approximation, 
which is usually valid under weak interaction. 
For the present, we go forward assuming that 
this description holds at least 
at the beginning of the contact.

To obtain the wave equation for the partial system $\varphi,$ 
we multiply (\ref{Schroedinger-eq}) by $\Psi^\ast$ from the left side, 
and integrate it over ${\bf R}$ 
under the condition of $\int d{\bf R}|\Psi|^2=1,$ 
where $d{\bf R}=d{\bf r}_{m+1}\cdots d{\bf r}_N.$
Then, we get 
\begin{eqnarray}\label{partial-sys}
[-i\hbar\frac{\partial}{\partial t}+\hat{h}-\lambda(t)]\varphi=0,\\
\hat{h}=\hat{h}_\varphi({\bf r})+V({\bf r}, t),
\end{eqnarray}
where
\begin{eqnarray}
\lambda(t)=\int^\infty_{-\infty}d{\bf R}\Psi^\ast({\bf R},t)[i\hbar\frac{\partial}{\partial t}-\hat{h}_\Psi({\bf R})]\Psi({\bf R},t).
\end{eqnarray}
and the `external field' $V({\bf r}, t)=\int d{\bf R}\hat{h}_{int}|\Psi|^2$.
If we combine equation (\ref{partial-sys}) with its complex conjugate, 
we get the equation for the current conservation and find $\lambda$ to be real. 
When $V$ is independent of $\varphi$ and $\lambda(t)=0$, 
equation (\ref{partial-sys}) is nothing but 
the Schr\"odinger equation for an isolated system. 
We, in principle, have to solve $\Psi$ and $\varphi$ at the same time.
In this sense, $V$ and $\lambda$ have 
non-linearity.%
\footnote{$V_\varphi$ and $\lambda_\varphi$ may be more appropriate notations.}
Although such non-linearity is a source of interest, 
for our purpose, it is more important 
that equation (\ref{partial-sys}) is a pseudo-linear form with 
the eigenvalue $\lambda$, i.e. 
\begin{eqnarray}\label{wave-eq}
[\hat{L}-\lambda(t)]\varphi=0,
\end{eqnarray}
where $\displaystyle\hat{L}=-i\hbar\frac{\partial}{\partial t}+\hat{h}.$
Therefore it is suggested that macroscopic states are classified 
under the `energy transfer' $\lambda(t)$, which we call {\it dissipation} (as is seen later in eq. (\ref{first-law}) in \S 3.2).
We notice that 
$\lambda$ has a nonlocal influence 
on the configuration space. 

Since $\Phi=\varphi\Psi$ is an approximate solution, 
the interaction makes $\displaystyle [-i\hbar\frac{\partial}{\partial t}+\hat{H}]\Phi\neq 0$
with the passage of time. 
In other words, 
the idea of `partial system' seems unstable. 
However, the standard quantum mechanics, 
which describes particles in an external field, teaches us 
the wide-ranging validity of this idea. 
In the next subsection we will investigate the origin of such stability 
of partial systems. 

\subsection{Emergence of a partial system}
If $\varphi$ is at a superposition state (in a secific basis), we naively predict 
that the interaction makes $\Psi$ move to the corresponding superposition state. 
This prediction, however, 
is sharply confronted with our daily experience 
consisting of macroscopic quantum systems. 
We see below how the occurrence of a superposition 
is suppressed by dissipation. 

We take up the state $\Phi_0$ again, 
which is one of the exact solution of the Schr\"odinger equation (\ref{Schroedinger-eq}).
Here we try to divide this system into two partial systems, $\varphi$ and $\Psi$, 
where we particularly consider the thermal contact of these systems. 
(Below, one can imagine $\Psi$ as a heat bath for $\varphi$.) 
Although $\Phi_0$ will be gradually away from 
the mean field solution $\Phi=\varphi\Psi,$ 
(i) we assume that we can expand it with the mean field solutions, $\Phi_\nu=\varphi_\nu\Psi_\nu$; 
\begin{eqnarray}\label{superposition}
\Phi_0=\sum_\nu\alpha_\nu\Phi_\nu, \\
\alpha_\nu=\int^\infty_{-\infty}d{\bf r}\int^\infty_{-\infty}d{\bf R}\Phi^\ast_\nu\Phi_0.
\end{eqnarray}
To obtain the description for the partial system, 
we multiply equation (\ref{Schroedinger-eq}) by $\sum_{\nu^\prime}\alpha^\ast_{\nu^\prime}\Psi^\ast_{\nu^\prime}$
from the left side and integrate it over ${\bf R}$. 
We get
\begin{eqnarray}\label{entangled}
\sum_\nu|\alpha_\nu|^2[\hat{L}-\lambda_\nu(t)]\varphi_\nu=0, \\
\hat{L}=-i\hbar\frac{\partial}{\partial t}+\hat{h}, 
\end{eqnarray}
where $\displaystyle\lambda_\nu(t)=\int d{\bf R}\Psi^\ast_\nu[i\hbar\frac{\partial}{\partial t}-\hat{h}_\Psi]\Psi_\nu$ and $\hat{h}=\hat{h}_\varphi+V_\nu$.
We have neglected the off-diagonal terms, 
$\displaystyle\int d{\bf R}\Psi^\ast_{\nu^\prime}\cdots\Psi_\nu$, 
which are small enough in most systems with large degrees of freedom. 

Now, to concentrate on the effect of the nonlocal term $\lambda,$ 
we neglect the non-linearity and the time dependence in $V.$
This is justified for the thermal contact. 
If we let $\phi_\nu$ be the solution for the isolated case 
(i.e. $\hat{L}\phi_\nu=0$), 
the solution for $[\hat{L}-\lambda_\nu]\varphi_\nu=0$ is expressed as 
\begin{eqnarray}
\varphi_\nu=\phi_\nu e^{i\Lambda_\nu(t)/\hbar},
\end{eqnarray}
where $\Lambda_\nu$ is the function that satisfies $\dot{\Lambda}_\nu=\lambda_\nu.$
Although we cannot exactly define the partial system at this point, 
from looking at equation (\ref{entangled}), 
it seems natural to define the partial wave function as 
\begin{eqnarray}\label{varphi}
\varphi=\sum_\nu|\alpha_\nu|^2 \phi_\nu e^{i\Lambda_\nu(t)/\hbar}.
\end{eqnarray}

We impose two more conditions on $\Lambda_\nu(t)$; 
(ii) the members of $\{\Lambda_\nu\}$ are dense enough, and then 
(iii) the absolute value of $\Lambda_\nu$ strongly depends on $\nu$ by the unit of $\hbar$. 
To satisfy the former assumption, it is necessary that the degrees of freedom of 
$\Phi_0$ is sufficiently large. 
The latter assumption is probably satisfied even in semi-macroscopic $\Phi_0$ 
because of the large factor 
$k_B/\hbar\simeq 1.3\times10^{11}$[K/s] for the thermal fluctuation. 
Then, $e^{i\Lambda_\nu/\hbar}$ (with the weight factor $|\alpha_\nu|^2$) 
strongly oscillates with $\nu$ 
and causes destructive interference among the states. 
Therefore the $\nu$'s whose $\Lambda_\nu(t)$ has an extreme value 
mainly contribute in equation (\ref{entangled}) and (\ref{varphi}).
Being the assumptions (i)-(iii) sufficiently satisfied, 
since only one $\nu_c$ survives, equation (\ref{entangled}) becomes 
\begin{eqnarray}\label{definite}
[\hat{L}-\lambda_{\nu_c}]\varphi_{\nu_c}=0,
\end{eqnarray}
and (\ref{varphi}) becomes 
\begin{eqnarray}
\varphi\to\varphi_{\nu_c}.
\end{eqnarray}
Similarly, 
we obtain $\Psi\to\Psi_{\nu_c}$
(the same index $\nu_c$ is chosen because 
$\lambda+\lambda_\Psi=const.$ for static $V$.)
Therefore we can regard the whole system as $\Phi=\varphi_{\nu_c}\Psi_{\nu_c}.$ 

Although the description $\Phi_0$ still holds, 
$\Phi$ is the right description to correspond 
to our recognition. 
In other words, the truly quantum description $\Phi_0$, 
whose degrees of freedom are mixed, is beyond 
our sense of reality. 
Even if such doubleness of the description is surprising, 
it becomes possible to make the concept of `partial system' coexist 
with the unitary time evolution of quantum mechanics.%
\footnote{%
Of course we can guess that $\Phi_0$ 
may be the partial system of a larger system, i.e. 
the isolation of $\Phi_0$ can be apparent (due to $\lambda_{\Phi_0}=0$). 
We, however, did not mention this in order to have our discussion converge. 
For the same reason, 
we did not discuss the case that 
the way to divide $\Phi_0$ into $\varphi$ and $\Psi$ is not unique.
}

Here we briefly comment on the measurement problem. 
For example, under strong Coulomb interaction 
one can expand the wave function 
with the coordinates at a specific time (according to Huygens' principle), 
and assign them indices $\nu$'s. 
(All of these localized states give extreme values to $\Lambda$, 
while the delocalized states disappear due to the interference.)
If only one $\nu$ survives along the above scenario, 
it is probable that this process corresponds to the position measurement. 
We stress that our scenario is deterministic 
in the description $\Phi_0$. 
Despite this, 
the destructive interference looks like acausal 
in the configuration space of the partial system $\varphi$. 
Only such nonlocal effect can 
describe the process $\varphi\to\varphi_{\nu_c}$ (see for example \cite{Bell1987}). 

We have to notice that $\Phi$ cannot be 
a permanent solution of any wave equation 
and equation (\ref{definite}) is also temporary. 
In other words, the quantum jump (transition) occurs corresponding to the change, 
$\nu_c\to\nu_{c^\prime}$.%
\footnote{If the energy is transferred only by real photons, 
$\Lambda(t)=n(t)\hbar$.}
Macroscopically, the transition, $\nu_c\to\nu_{c^\prime}
\to\cdots\to\nu_{c^{\prime\prime}}$, causes 
`thermal' energy transfer and `friction'. 

There is no denying the possibility that 
a superposition of different $\nu$'s survives when 
the assumptions (i)-(iii) are insufficiently satisfied. 
In common macroscopic objects, however, 
we can safely expect that 
the thermal fluctuation is always suppressing the quantum fluctuation. 
Therefore we assume 
$\Phi=\varphi\Psi$ in the next section.

\section{Action and physical quantities}
While we introduced 
$\lambda(t)$ by integrating 
the external degrees of freedom in \S 2, 
we here present the discussion 
in a more deductive manner. 
We study the conservation laws for macroscopic systems 
and verify the consistency of our approach. 

From our daily experience, 
a macroscopic object, which justly consists of quanta, occupies 
a certain domain in space and time. 
Therefore it occurs to us that 
from a certain scalar, namely an action, 
we derive physical quantities 
as well as in classical mechanics. 
These quantities correspond to the expectation values in quantum mechanics, 
and the action should also produce the corresponding wave equation. 

Although the friction discussed in \S 2.2 is important, 
we here concentrate on the quasistatic case, i.e.  
considering the term in which the wave equation holds. 

\subsection{Action principle}
From the following action%
\footnote{%
The action is also expressed as $S=\Lambda(t)\bigr|^{t_2}_{t_1}$, where $\Lambda$ is the time integrated dissipation in \S 2. }, 
we can develop all of the discussion. 
\begin{eqnarray}\label{action}
S=\int^{t_2}_{t_2} dt\int^\infty_{-\infty} d{\bf r}\varphi^\ast \hat{L} \varphi, \\
\hat{L}=-i\hbar\frac{\partial}{\partial t}+\hat{h},
\end{eqnarray}
where 
\begin{eqnarray}
\hat{h}=\hat{h}_\varphi({\bf r})+V({\bf r},t).
\end{eqnarray}
Under the norm condition $\displaystyle\int^\infty_{-\infty} d{\bf r}|\varphi|^2=1$, 
we require this action to satisfy the stationary condition 
in the $3m+1$ dimensional configuration space-time. 
For example, with $\delta\varphi^\ast(t_1)=\delta\varphi^\ast(t_2)=0$, 
the variation $\varphi^\ast+\delta\varphi^\ast$
gives the Schr\"odinger equation containing 
Lagrange multiplier $\lambda(t)$;
\begin{eqnarray}
\frac{\delta}{\delta \varphi^\ast}\left[S-\int^{t_2}_{t_2}dt\lambda(t)\int^\infty_{-\infty} d{\bf r} |\varphi|^2\right]=[\hat{L}-\lambda(t)]\varphi=0.
\end{eqnarray}
This wave equation is the same as equation (\ref{wave-eq}) in \S2 
and includes the isolated case as $\lambda(t)=const.$
(Being $V$ and $\lambda$ real, the variation $\delta {\rm arg}\varphi$ 
gives the current conservation.)

It should be justified to ignore 
the non-linearity of $V$ and $\lambda$ in the 
above variation.
To give a suggestion about this, 
we consider the total (mean field) action; 
\begin{eqnarray}
\mathcal{S}&=&\int dtd{\bf r}d{\bf R}\Phi^\ast[-i\hbar\frac{\partial}{\partial t}+\hat{H}]\Phi\\
&=&S+\int dtd{\bf R}\Psi^\ast[-i\hbar\frac{\partial}{\partial t}+\hat{h}_\Psi]\Psi.
\end{eqnarray}
We assign $\varphi_\alpha$ to 
$\varphi$ and $\varphi^\ast$, 
as well as $\Psi_\beta$ to $\Psi$ and $\Psi^\ast$, for convenience. 
The variational parameters $\delta\varphi_\alpha$ and  
$\delta\Psi_\beta$ give four wave equations. 
Firstly we solve the equations $\delta\mathcal{S}/\delta\Psi_\beta=0$ 
and obtain $\Psi_\beta$ as functionals of $\varphi_\alpha$.
Substituting this, we obtain $\tilde{\mathcal{S}}$ 
as an {\it implicit} functional of $\varphi_\alpha$. 
Let us express such variation as $d\varphi_\alpha$. 
Then the variation of this action is represented as 
\begin{eqnarray}
\frac{d\tilde{\mathcal{S}}}{d\varphi_\alpha}=\frac{\delta S}{\delta\varphi_\alpha}+\frac{\delta\tilde{\mathcal{S}}}{\delta \Psi_\beta}\frac{d\Psi_\beta}{d\varphi_\alpha}=0.
\end{eqnarray}
We need to consider only the first term because the second term is zero.

\subsection{Physical quantities and classical action}
We know that the conservation laws and the symmetry of a system 
are closely connecting each other; symmetry operations guide to 
the relation between a wave function and physical quantities. 
To simplify the expression 
we show the case of one-body momentum and energy below.

It is obvious that the action doesn't change when 
the integral parameter $x$ is shifted to $x+\Delta x$. 
Due to this shift, there are two types of change in the integral;
one is caused from 
$\displaystyle\Delta\varphi=\frac{\partial\varphi}{\partial x}\Delta x$ and the other is 
from $\displaystyle\Delta V=\frac{\partial V}{\partial x}\Delta x$.
We represent these contributions such as $\displaystyle\Delta S=\Delta_\varphi S+\Delta_V S$.
Using the wave equation, the first term is transformed into 
$\displaystyle\Delta_\varphi S=\int^\infty_{-\infty}d{\bf r}\varphi^\ast\frac{\hbar}{i}\frac{\partial}{\partial x}\varphi\biggr|^{t_2}_{t_1}\Delta x$, 
which is related to the conservation law because
only the initial and the final state 
contribute to this. 
Of course this is what the orthodox quantum mechanics teaches us 
as the $x$ component of the momentum. 
Therefore, if we use the notation 
$\displaystyle p_x(t)=\int^\infty_{-\infty}d{\bf r}\varphi^\ast\frac{\hbar}{i}\frac{\partial}{\partial x}\varphi$, we find 
\begin{eqnarray}
\frac{\Delta S}{\Delta x}=p_x\Bigr|^{t_2}_{t_1}+\int^{t_2}_{t_1} dt\int^\infty_{-\infty}d{\bf r}|\varphi|^2\frac{\partial}{\partial x}V=0,\\
p_x\Bigr|^{t_2}_{t_1}=p_x(t_2)-p_x(t_1).
\end{eqnarray}
The time derivative of this equation gives Newton's second law; 
\begin{eqnarray}\label{Newton}
\dot{{\bf p}}=-\int^\infty_{-\infty} d{\bf r}|\varphi|^2\nabla V.
\end{eqnarray}

Similarly, starting from $\displaystyle\frac{\Delta S}{\Delta t}=\dot{\Lambda}(t)\Bigr|^{t_2}_{t_1}=\lambda(t)\Bigr|^{t_2}_{t_1}$, we obtain the conservation law for the energy; 
\begin{eqnarray}\label{first-law}
\mathcal{E}\Bigr|^{t_2}_{t_1}=-\lambda(t)\Bigr|^{t_2}_{t_1}+\int^{t_2}_{t_1} dt\int^\infty_{-\infty}d{\bf r}|\varphi|^2\frac{\partial}{\partial t}V,\\
\mathcal{E}\equiv\int^\infty_{-\infty} d{\bf r}\varphi^\ast i\hbar\frac{\partial}{\partial t}\varphi.
\end{eqnarray}
This corresponds to the first law of thermodynamics. 
$\lambda(t)\Bigr|^{t_2}_{t_1}$ is understood as the amount of heat 
to flow out. 

These results are easily extended for $N$-body case. 
Finally we consider the case that 
the system can be described within 
the framework of classical mechanics 
by eliminating the degrees of freedom. 
Let us assume that the system consists of same kind of particles and 
pay attention only to the center-of-mass motion. 
Considering the translation of all the elements 
at the same time, we obtain the variation of the action as%
\footnote{$\hat{h}_\varphi$ is invariant under this transformation.}
\begin{eqnarray}
\delta S&&=\Delta_\varphi S\\&&={\bf p}\cdot\Delta {\bf q}-\mathcal{E}\Delta t,
\end{eqnarray}
where 
$\displaystyle {\bf p}=<\sum_i\hat{\bf p}_i>=\int d{\bf r}\varphi^\ast \frac{\hbar}{i}\sum_i\nabla_i\varphi$.
Here we can interpret $(\Delta{\bf q}, \Delta t)$ as the infinitesimal path 
of the center-of-mass. 
We also assume that the internal state does not change; $\lambda$ 
and the dispersion $<\sum_i\hat{\bf p}_i^2>/N-<\sum_i\hat{\bf p}_i>^2/N^2$ are constant.
Moreover, using the approximation $\int d{\bf r}|\varphi|^2V\simeq V({\bf q})$, 
we obtain 
$\displaystyle\mathcal{E}({\bf p},{\bf q})=\frac{{\bf p}^2}{2M}+V({\bf q})+const.$, 
where $M=Nm$ is the total mass. 
There remains only two degrees of freedom, i.e. the path and the momentum. 
Then we obtain the classical action; 
\begin{eqnarray}
S_{path}({\bf p},{\bf q})&&=\int\Delta_\varphi S\\
&&=\int^{t_2}_{t_1} ({\bf p}\cdot\dot{\bf q}-\mathcal{E}({\bf p},{\bf q}))\Delta t.
\end{eqnarray}
$\delta S_{path}/\delta{\bf p}=0$ and $\delta S_{path}/\delta{\bf q}=0$ 
give ${\bf p}=M\dot{\bf q}$ and 
equation (\ref{Newton}), respectively.

\section{Conclusions}
We have studied the connection 
between the classicality and 
the dissipation in order to derive the macroscopic reality 
from quantum mechanics. 
We presented the scenario, where the dissipation causes 
the destructive interference between superposition states. 
In other words, in this case, the macroscopic system 
can be described as a aggregate of stable partial systems and 
it behaves like a classical system.

While we can understand a wave function 
as a matter wave in the configuration space (\S 3), 
the nonlocality of $\lambda$ is also important 
when we consider the dynamics of the system (\S 2). 
This suggests the totality (namely, the imperfection of 
the `partial system'), which has been often mentioned 
(for example in \cite{Wheeler1983}) in the context of the measurement problem. 
As the result, the macroscopic reality consists of the processes (expressed by $\Phi$)
that make the action (namely, the time integrated dissipation) 
take an extreme value. 
Behind this, the unknowable linear description ($\Phi_0$) exists.

If it is difficult, there is a possibility 
of checking the correctness of our scenario 
to use so-called `macroscopic quantum systems', i.e. 
the collapse of a superposition 
can be observed by controlling the assumptions (ii) and (iii) in \S 2.2.
The observation of such collapse has already been discussed 
(for example in \cite{Leggett2002}).
Of course, by numerical calculation with proper approximation, 
we can confirm the validity of the assumptions (i)-(iii). 

Although we found that there are two types of time-evolution in our macroscopic reality, 
i.e. the transition $\varphi_{\nu_c}\to\varphi_{\nu_{c^\prime}}$ 
and the linear-time dependent evolution, 
the connection between the former time-asymmetric process and 
the second law of thermodynamics must be studied. 
We recognize lack of quantitative evaluation 
and applications for concrete cases throughout this study.  
Critical examination of this rough sketch is necessary.

\end{document}